\documentclass{article}

\usepackage{physics}
\usepackage{braket}
\usepackage[square,sort,comma,numbers]{natbib}
\usepackage[english]{babel}
\usepackage{graphicx}
\usepackage{float}
\usepackage{titling}
\usepackage{lipsum}
\usepackage{indentfirst}
\usepackage{amsmath}
\usepackage{amssymb}
\usepackage{caption}
\usepackage{sidecap}
\usepackage{mathrsfs}
\usepackage[a4paper, total={5.75in, 9in}]{geometry}

\sidecaptionvpos{figure}{c}

\title{The arrow of time and a-priori probabilities}
\author{Sivapalan Chelvaniththilan \\ email: niththilan@gmail.com}
\date{\today}
\begin{document}
\maketitle
\begin{abstract}
The second law of thermodynamics is asymmetric with respect to time as it says that the entropy of the universe must have been lower in the past and will be higher in the future. How this time-asymmetric law arises from the time-symmetric equations of motion has been the subject of extensive discussion in the scientific literature. The currently accepted resolution of the problem is to assume that the universe began in a low entropy state for an unknown reason. But the probability of this happening by chance is exceedingly small, if all microstates are assigned equal a-priori probabilities. In this paper, I explore another possible explanation, which is that our observations of the time-asymmetric increase of entropy could simply be the result of the way we assign a-priori probabilities differently to past and future events.
\end{abstract}

\section{Introduction}
The past and the future are different from each other. This is something we all know from our everyday experience. We can remember the past but not the future. A blob of diffusing particles is likely to become more spread out in the future and likely to have started out in a more concentrated region in the past. A flower on a plant would have been a bud in the past and may become a fruit or wither away in the future.  \\

But proving such differences mathematically has been difficult, since the equations of motion are symmetric under time-reversal. This means, for example that if we imagine some ordinary occurrence happening in reverse, such as broken pieces of a glass object spontaneously flying up from the ground and joining together to form the unbroken object, this would be perfectly consistent with the laws of motion. Yet, we never observe anything like this happening. \\
 
A partial explanation for these differences is that a system would always tend to go to high entropy macrostates, i.e. macrostates that are highly disordered, simply because there are far more ways for a system to be disordered than for it to be ordered. (A microstate is defined as an exact state of a system with the configurations of every constituent particle specified exactly. In this paper, I define two microstates as belonging to the same macrostate if an observer cannot distinguish between them through measurements on the whole system.) \\

Since the entropy S of a macrostate with W microstates in it is defined as $S=k_B\ln W$, a macrostate with even a slightly higher entropy would have an extremely larger number of microstates in it. For example, when the entropy increases by $0.01JK^{-1}$, the number of microstates increases by a factor of $10\wedge(3\times 10\wedge20)$. This number is 1 followed by 300 million trillion zeros. So any random change from one microstate to another will almost certainly be a change to a higher entropy macrostate, and this is why entropy always increases. \\

But Boltzmann \cite{boltzmann1896entgegnung} (and later Feynmann \cite{feynman1965feynman} and Eddington \cite{eddington1931end}) have pointed out a major incompleteness in the above reasoning, which can be summarised as follows: A system is more likely to go to in a higher entropy macrostate in the future because such a macrostate has many more microstates. Boltzmann asks us what would happen if we apply the same reasoning to the past rather than the future. Should we conclude that the system is more likely to have been in a higher entropy macrostate in the past for because such a macrostate has many more microstates? This would contradict all our observations of systems having lower entropy in the past and yet this is the conclusion we would reach of we apply the reasoning in a time-symmetric way. \\
 
One possible way to explain this is to simply assume that the universe began in some low entropy state. Then even though the laws of motion are time-symmetric, the boundary conditions with which these laws are applied are time-asymmetric (with a low entropy condition in the past and not in the future). This time-asymmetry of the boundary condition can be thought of as providing an explanation for the time-asymmetry in our observations. But there is a severe drawback in explaining the time-asymmetry in this way. This becomes apparent when we ask ourselves why the universe started out in this low entropy macrostate and not a higher one. Was it just by random chance? But using $S=k_B\ln W$ again, the probability of a random selection resulting in a low entropy macrostate is neglegibly small. \\

\begin{SCfigure}[][h!]
\centering
\includegraphics[scale=0.75]{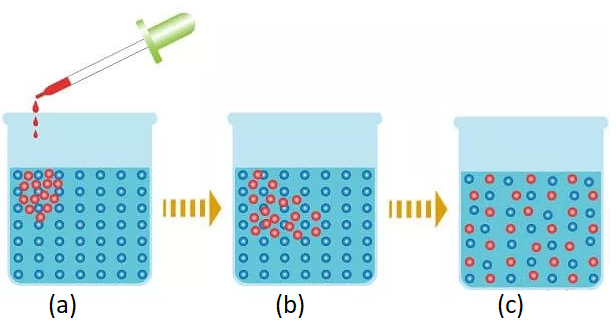} 
\caption{A drop of dye spreading out due to diffusion. Figure copied from \cite{biologydictionary:2020} and modified}
\label{fig:1}
\end{SCfigure}

In order to understand this time-asymmetry more clearly, we need to look at the assumption behind these calculations, namely that all microstates have equal a-priori probabilities. This is usually assumed in calculations because it seems to be the most natural way to assign a-priori probabilities. (Giving higher a-priori probabilities to some microstates than to others would be quite arbitrary). This assumption gives the correct results (in agreement with observations) when calculating the state of a system at a later time using the state at an earlier time. For example, consider the diffusion of a dye in water as shown in Figure \ref{fig:1}. \\

 Figure \ref{fig:1}a shows a drop of dye being placed in a beaker of water. \ref{fig:1}b shows the dye slightly spread out after some time and \ref{fig:1}c shows it fully spread out after a longer time. Suppose we run a simulation of this system for a certain duration forward in time starting from the state in Figure \ref{fig:1}b. Suppose we do this using the assumption of equal a-priori probabilities for microstates. The probability that this simulation would reach a state like the one in Figure \ref{fig:1}c is quite high because such a macrostate contains a much higher number of microstates than in Figures \ref{fig:1}a and \ref{fig:1}b. Thus the result of a simulation based on this assumption agrees with observation. \\

Does this assumption give the correct result when doing the calculation the other way, that is, when calculating the state of the system at an earlier time given the state at a later time? For example, suppose we again start the simulation from  Figure \ref{fig:1}b and run it backward in time. Would this simulation correctly predict the dye would be more concentrated at the earlier time? That is, would it correctly predict the state shown in Figure \ref{fig:1}a? Because of the time-symmetry of the diffusion equation, this simulation would be exactly the same as the previous one. Therefore, the result is far more likely to be like the one in Figure \ref{fig:1}c than Figure \ref{fig:1}a. We are forced to conclude that the law of equal a-priori probabilities for microstates is not appropriate for calculations of earlier states from later states. \\

There is a possible objection to this reasoning. Even though just the system under consideration (the water and the dye) might have a higher probability of evolving from the state in \ref{fig:1}b to the state in \ref{fig:1}c when it is simulated backwards, we need to consider both the system and its surroundings. The surroundings might contain conscious observers - persons - who might have memories of what the system was like at an an earlier time. These memories would be consistent with only Figure \ref{fig:1}a and not \ref{fig:1}c since the observers would have seen the dye spreading out from the initial location (1a $\rightarrow$ 1b) and not spontaneously separating from the water and getting concentrated at a certain location (1c $\rightarrow$ 1b) \\

How do we reconcile this fact about memories with the time-symmetry of the equations of motion? This is the problem explored by Boltzmann \cite{boltzmann1896entgegnung}, Feynmann \cite{feynman1965feynman} and Eddington \cite{eddington1931end}. The essence of their conclusion is that such a hypothetical simulation (involving both the system and observers) when run backwards would still result in the dye being more spread out in the past, just as it would be in the future. This means that the principle of equal a-priori probabilities implies that the observers' memories have a high probability of being false memories. (i.e. the simulation would predict that the dye was more spread out in the past while the observers remember seeing it concentrated in the past). \\

But the above reasoning applies not only to memories of watching a drop of dye diffuse. It applies to all memories - since our memories are of low entropy macrostates in the past and low entropy macrostates have fewer microstates, the principle of equal a-priori probabilities implies that all these memories have a high probability of being false. This principle leads to the conclusion that any given observer has a high probability of being just an isolated brain (since the memories of the rest of the body could also be false) with a fleeting existence in a chaotic high entropy universe, and having false memories of being in an ordered, low entropy universe filled with other living beings. Since this conclusion goes so strongly against our observations, it is called the Boltzmann Brain paradox \cite{feynman1965feynman}. \\

The usual way of avoiding this paradox, as explained earlier, is to assume that the universe began in a low entropy macrostate. But this combined with the principle of equal a-priori probabilities makes the initial macrostate of the universe a very unlikely one. The purpose of this paper is to suggest a different way of avoiding the paradox - by changing the assumption about a-priori probabilities. Specifically, the suggestion is that assuming that the universe had equal probabilities of starting out in each macrostate - rather than microstate - would lead to conclusions that are more consistent with everyday observations instead of leading to the Boltzmann Brain paradox. \\

But this new assumption seems very unnatural. The microstates of a system are fundamental to that system but macrostates have more to do with our observations of the system.(Since I am assuming that two microstates are part of the same macrostate if they lead to the same observations, this means that in the example of the dye, two microstates which have the same distributions  of the concentration of the dye but differ in the exact positions of the dye and water molecules will be in the same macrostate). \\

In Sections 2 to 4 below I will show how this new assumption is consistent with our observations of thermodynamic phenomenon such as diffusion and heat flow. Then in Section 5, I will explore the possible reasons why this seemingly unnatural choice of a-priori probabilities might perhaps be more natural than equal a-priori probabilities for microstates. It must also noted that this new assumption only assigns equal a-priori probabilities to macrostates only at the beginning of the universe (i.e. the universe is equally likely to have started out in each macrostate). The probabilities of each microstate or macrostate at later times must be calculated using these a-priori probabilities and the transition probabilities (i.e. the probabilities for a particular initial state to evolve into another state at a later time). These transition probabilities are determined by the equations of motion. 

\section{Applying the new assumption to a few examples}
Since the new assumption is that the universe had equal a-priori probabilities of starting out in each macrostate, in order to check if it gives correct predictions, we would need to consider a system that is complex enough to be a model of the whole universe, with conscious observers in it. Since this is a difficult task, I will not attempt it. Instead I will consider very simple systems that evolve according to a fixed set of rules. I will specify which properties of the system can be measured by an observer, but I will not include any mathematical model of the memory or cognitive process of the observer. So even though the calculations below hint that the assumption might be correct, they do not offer a definite proof.

\subsection{A system of particles that can be in two possible states}
Consider an set of 100 particles, each of which can be in one of two states which I will call ``up'' and ``down''. Let us also imagine an observer who can measure the number M of particles that are in the up state (this number can take 101 values from 0 to 100) but cannot find out which particles are up and which ones are down. Hence this system has $2^{100}$ microstates which can be classified into 101 macrostates. The number of microstates in the macrostate labelled by the integer M is given by

\begin{eqnarray}
\frac{100!}{M!(100-M)!}
\label{eqn:1}
\end{eqnarray}

i.e., it is equal to the number of ways of selecting M particles from 100 particles. So the M=100 macrostate has only one microstate as all the particles must be up. The M=99 one has 100 microstates because one particle must be down and it could be any one of the 100 particles. The M=98 one has 4950 microstates as 2 particles must be down and there are 4950 possible ways to select 2 particles from 100 and so on. The M=50 macrostate has the highest number of microstates, followed by the ones near M=50. \\

Let us model the time evolution of this system as follows. At discrete time steps, one of the 100 particles is chosen at random, and its state is flipped i.e., if it was up it is changed to down or vice versa. Suppose the system is in the M=99 macrostate at a particular time $t_0$. Then when a random particle is flipped, it has a probability of 0.99 of going to the M=98 state and a probability of 0.01 of going to the M=100 state at $t_0+\delta t$. This is because if any of the 99 particles in the up state were flipped the system would go to the M=98 macrostate and if the one particle in the down state were flipped, the system would go to the M=100 macrostate. \\

Since the M=98 macrostate has more microstates (and hence more entropy) than the M=99 one which in turn has more than the M=100 one, the above reasoning correctly predicts that the system is more likely to transition to a higher entropy macrostate at $t_0+\delta t$. But what if we use the same reasoning to find the probabilities of the states at $t_0-\delta t$? Because of the time-symmetric evolution, we get the same probabilities (0.01 for M=100 and 0.99 for M=98) and hence we reach the conclusion that entropy must have been higher at $t_0-\delta t$. This is contrary to our everyday experience because just as in the case of the dye, if we saw the system at M=99, our intuitive guess would be that it must have been at M=100 earlier (as this macrostate is the most likely to evolve into the current macrostate). \\

Could the reason for the non-intuitive conclusion from the calculations be an implicit assumption about a-priori probabilities? To find out, let us do the same calculation using Bayes' theorem. We need to consider the probabilities of different macrostates at $t_0-\delta t$ transitioning to M=99 at $t_0$. Here, the only two possiblilities are M=98 and M=100. In order to apply Bayes theorem, we simply multiply the a-priori probability for each one by the probability for it to transition to M=99 and then divide by the total.

\begin{eqnarray}
P(M=98 ~\mathrm{at}~ t=t_o-\delta t)=\frac{P_a(98)P(98\rightarrow 99)}{P_a(98)P(98\rightarrow 99)+P_a(100)P(100\rightarrow 99)}  \nonumber \\ \nonumber \\ 
=\frac{  \left( \scalebox{1.5}{$\frac{4950}{2^{100}}$} \right) \times 0.02}{ \left( \scalebox{1.5}{$\frac{4950}{2^{100}}$} \right) \times 0.02+ \left( \scalebox{1.5}{$\frac{1}{2^{100}}$} \right) \times 1}=0.99 ~~~~~~~~~~~~~~~~~~~~
\label{eqn:2}
\end{eqnarray}

Here, I denote by $P_a(98)$ and $P_a(100)$ the a-priori probabilities of the M=98 and M=100 macrostates respectively. (When all microstates are given equal a-priori probabilities, the a-priori probability of a macrostate is simply the number of microstates in it divided by the total number of microstates, which in this case is $2^{100}$.) I denote by $P(98\rightarrow 99)$ the probability of a random microstate in the M=98 macrostate to transition to the M=99  macrostate. This probability is 2/100 because for this to happen, a particle in the down state must be flipped to the up state and only 2 of the 100 particles are in the down state.  \\

Similarly the probability for it to have been in the M=100 macrostate is 

\begin{eqnarray}
P(M=100 ~\mathrm{at}~ t=t_0-\delta t)=\frac{P_a(100)P(100\rightarrow 99)}{P_a(98)P(98\rightarrow 99)+P_a(100)P(100\rightarrow 99)}  \nonumber \\ \nonumber \\ 
=\frac{ \left( \scalebox{1.5}{$\frac{1}{2^{100}}$} \right) \times 1}{ \left( \scalebox{1.5}{$\frac{4950}{2^{100}}$} \right) \times 0.02+ \left(\scalebox{1.5}{$\frac{1}{2^{100}}$} \right) \times 1}=0.01~~~~~~~~~~~~~~~~~~~~
\label{eqn:3}
\end{eqnarray}

So the calculation using Bayes' theorem gives the same (non-intuitive) result as the direct calculation only if we give equal a-priori probabilities to microstates. What if we instead assign equal a-priori probabilities to macrostates, that is, if we consider each of the 101 macrostates to have the same a-priori probability of 1/101? Then the above results would change to

\begin{eqnarray}
P'(M=98 ~\mathrm{at}~ t=t_o-\delta t)=\frac{(1/101)\times 0.02}{(1/101)\times 0.02+(1/101)\times 1}=\frac{1}{51} \nonumber \\ \nonumber \\ 
P'(M=100 ~\mathrm{at}~ t=t_o-\delta t)=\frac{(1/101)\times 1}{(1/101)\times 0.02+(1/101)\times 1}=\frac{50}{51}
\label{eqn:4}
\end{eqnarray}

Now the prediction is that the system was initially more likely to have been in M=100. So if we assign equal a-priori probabilities to macrostates when calculating probabilities of earlier events we get results that are more consistent with our intuitive sense of probabilities.

\subsection{Diffusion and heat flow: a qualitative explanation}
Now let us apply the same reasoning to the spreading of a drop of dye in water. Suppose we see the dye as it is in Figure \ref{fig:1}b. Our intuitive guess would be that at an earlier time it must have been in a state similar to \ref{fig:1}a. This is because \ref{fig:1}a transitioning to \ref{fig:1}b is far more likely than \ref{fig:1}c transitioning to \ref{fig:1}b. When the dye is as in \ref{fig:1}a, almost any movement of the dye molecules will result in a state such as \ref{fig:1}b. Hence the transition probability P(1a $\rightarrow$ 1b) is high. But when the dye is in a state such as \ref{fig:1}c, it would be a rare coincidence if all the molecules moved in such a way as to make the dye more concentrated, as in \ref{fig:1}b. Hence the transition probability P(1c $\rightarrow$ 1b) is low. \\

Our guess that the past state was \ref{fig:1}a would be confirmed if we ask anyone who has seen the dye earlier. But the calculation using the equations of motion, being time-symmetric, predict (wrongly) that it is more likely to have been like \ref{fig:1}c earlier, just as it is more likely  to go to such a state in the future. Why does it make such a prediction? \\

\begin{SCfigure}[][h!]
\centering
\includegraphics[scale=0.65]{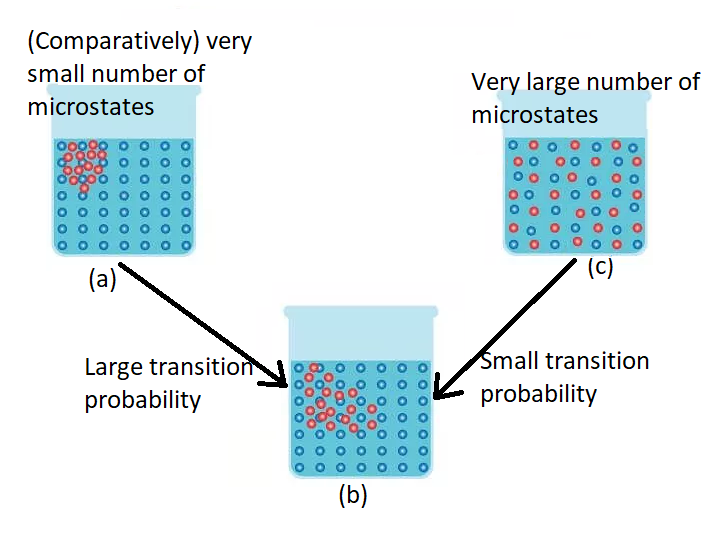} 
\caption{A simple explanation for why the diffusion calculation with equal a-priori probabilities for microstates gives a non-intuitive result when calculating an earlier state from a later state.}
\label{fig:2}
\end{SCfigure}

This is because the probability of the past macrostate, when calculated using Bayes' theorem, is proporional to both the a-priori probability and the transition probability to the current state. The macrostate in \ref{fig:1}c has many more microstates (and hence a much higher a-priori probability) than \ref{fig:1}a and the effect of this outweighs the effect of the smaller transition probabilities.  This is summarised in Figure \ref{fig:2}.\\

So how do we prevent this contradiction between our intuition and the calculation? If we assign equal a-priori probabilities to macrostates instead, then \ref{fig:1}a and \ref{fig:1}c would have the same a-priori probabilities even though the number of microstates in them are different. Then, the calculation would predict that \ref{fig:1}a is more likely, in agreement with our intuition. \\

Note that this is not a fully accurate example of the use of our new assumption. The new assumption is that all macrostates have equal a-priori probabilities at the beginning of the universe, not at some arbitrary time in the past. But since an accurate use of this assumption would require a model of the whole universe, I have instead given a simpler but inaccurate application of it, just to give a basic idea of how it works. \\

The above example can also be more clearly understood by comparing it to the calculation on the system of 100 particles done in Section 2.1. The M=100, M=99 and M=98 macrostates in the above calculation are analogous to the dye in Figures \ref{fig:1}a, \ref{fig:1}b and \ref{fig:2}c respectively. The transition probability $P(98\rightarrow 99)$ is smaller than $P(100\rightarrow 99)$  but this is outweighed by the higher number of microstates in M=98.  \\

\begin{SCfigure}[][h!]
\centering
\includegraphics[scale=0.6]{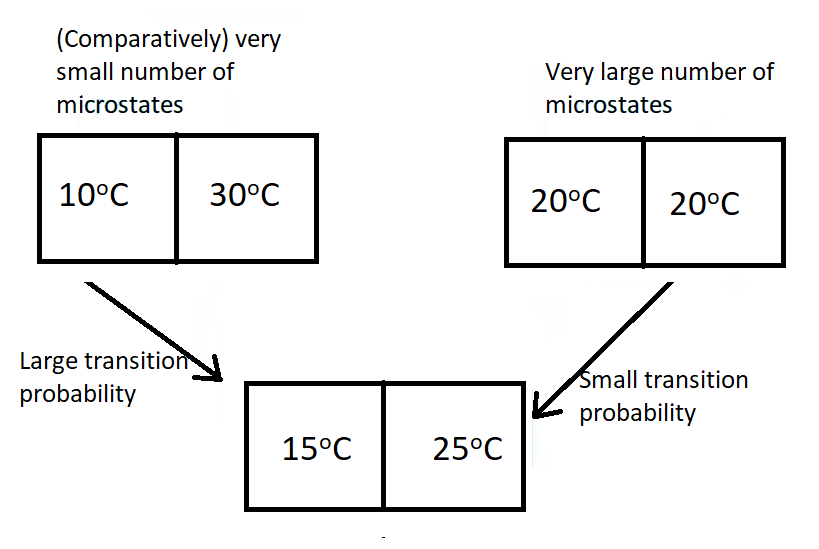} 
\caption{A similar explanation for heat flow.}
\label{fig:3}
\end{SCfigure}

A similar reasoning can be used to explain any time-asymmetric thermodynamic process, such as heat flow, which is illustrated in Figure \ref{fig:3}. If we see two objects in thermal contact, at two different temperatures, say $15^\circ C$ and $25^\circ C$ respectively, we will deduce that (if they have been in thermal contact for some time), their temperature difference would have been higher earlier (say $10^\circ C$ and $30^\circ C$). But a calculation using equal a-priori probabilities for microstates contradicts our intuitive guess and predicts (wrongly) that the system is more likely to have been in a state with equal temperatures. Again this conclusion can be avoided by giving equal a-priori probabilities to macrostates.

\section{More detailed calculations}
Below, I will try to provide more detailed calculations to complement the qualitative arguments in the previous section.. Let us study the diffusion of a gas of identical particles in one dimension. Let us model this problem by dividing the one dimensional line into segments of width $\epsilon$. Let $\delta$ be the size of one particle, so that the maximum number of particles in each segment is $\epsilon/\delta$. The observer can measure the number $n_i$ of particles in each segment i but cannot obtain more accurate knowledge of their locations. So the macrostates of this system will be labeled by a series of numbers $(\cdots,n_{-1}, n_0, n_1, n_2, \cdots)$ which denote the number of particles in each segment.. This can be thought of as the density distribution of the diffusing particles. Now let us calculate the number of microstates in each macrostate. \\

Consider the segment 1. Since the maximum number of particles in it is $\epsilon/\delta$, it can be thought of as having $\epsilon/\delta$ ``slots'' in which particles can be filled. Out of these, $n_1$ are filled and $\epsilon/\delta-n_1$ are empty. The number of possible ways in which this can happen is

\begin{eqnarray}
\frac{(\epsilon/\delta)!}{n_1!(\epsilon/\delta-n_1)!}\approx\frac{(\epsilon/\delta)^{n_1}}{n_1!}
\label{eqn:5}
\end{eqnarray}

where I have made an approximation by assuming that $\epsilon/\delta \gg n_1$. So the total number of microstates of the system is 

\begin{eqnarray}
\cdots \frac{(\epsilon/\delta)^{n_{-1}}}{n_{-1}!}  \frac{(\epsilon/\delta)^{n_0}}{n_0!}  \frac{(\epsilon/\delta)^{n_1}}{n_1!}  \frac{(\epsilon/\delta)^{n_2}}{n_2!} \cdots = \frac{(\epsilon/\delta)^N}{\cdots n_{-1}!n_0!n_1!n_2!}=(\epsilon/\delta)^N\prod_i\frac{1}{n_i!}
\label{eqn:6}
\end{eqnarray}

where $N=\sum_i n_i$ is the total number of particles in the diffusing gas. Now let us model the diffusion process as follows: At each time step one of the N particles is selected at random and is moved to a random slot in either the next or the previous segment with equal probability. Let us consider the how the distribution will change after K such steps. I assume that $K\ll N$ so that the change to the distribution of the particles is small. I also assume that $K\gg 1$ which will enable us to use the Sterling approximation \cite{dutka1991early} in the calculations below. Out of these K steps, let $R_i$ be the number of steps in which a particle from segment i moved to the right (i.e. moved to segment i+1) and $L_i$ be the number of steps in which a particle from segment i moved to the left (to segment i-1). Since one particle is moved in each step, this means $\sum_i(R_i+L_i)=K$.\\ The probability for $R_i$ and $L_i$ to take a set of particular values is given by

\begin{eqnarray}
P(R_i,L_i)=K!~ \prod_i  \left(\frac{n_i}{2N}\right)^{R_i+L_i} \frac{1}{ R_i!L_i! }
\label{eqn:7}
\end{eqnarray}

This is because $(n_i/N)$ is the probability for the particle selected in a particular step to be in segment i and the probability for it to be moved either left or right is $1/2$. Multiplying these factors for each segment i gives $\prod_i (n_i/2N)^{R_i+L_i}$. This is then multiplied by $K!/\prod_i(R_i!L_i!)$ which (from basic combinatorics) is needed because the values of $R_i$ and $L_i$ do not depend on the order of the steps. Using the Sterling approximation, this becomes

\begin{eqnarray}
\ln P(R_i,L_i)=K\ln K+ \sum_i  \bigg[ (R_i+L_i)\ln \left(\frac{n_i}{2N}\right)-R_i\ln R_i -L_i \ln L_i \bigg]
\label{eqn:8}
\end{eqnarray}

Using the method of Lagrange multipliers \cite{paulsonlinenotes:2019} we can find the maximum value of this quantity subject to the constraint $\sum_i (R_i+L_i)=K$. It will be maximum when

\begin{eqnarray}
R_i=L_i=\frac{Kn_i}{2N}
\label{eqn:9}
\end{eqnarray}

Now the change in $n_i$ after K steps (which can be expressed as $K\tau \partial n/\partial t$ where $\tau$ is the time interval between consecutive steps) is $R_{i-1}+L_{i+1}-R_i-L_i$ because $R_{i-1}+L_{i+1}$ is the number of particles that enter segment i and $R_i+L_i$ is the number of particles that leave it. When $R_i$ and $L_i$ take the maximum probability values found above, this quantity becomes

\begin{eqnarray}
R_{i-1}+L_{i+1}-R_i-L_i=\frac{K}{2N}(n_{i-1}+n_{i+1}-2n_i)
\label{eqn:10}
\end{eqnarray}

But in the continuum limit, this is $(K\epsilon^2/2N)\partial^2 n/\partial x^2$. This means we have derived the diffusion equation 

\begin{eqnarray}
\frac{\partial n}{\partial t}= \frac{\epsilon^2}{2N\tau} \frac{\partial^2 n}{\partial x^2}
\label{eqn:11}
\end{eqnarray}

But in the above derivation, we have assumed the initial state (at a particular time, say, $t_0$) and derived the most probable state at a later time (after K steps). So we have proved the diffusion equation only for $t>t_0$. For $t<t_0$, we can use the same steps as above, the only difference being that the change in the density distribution K steps before $t_0$ is now equal to $-K\tau \partial n/\partial t$ since we are considering the possible states K steps \emph{before} $t_0$. Thus we reach the incorrect conclusion that

\begin{eqnarray}
\frac{\partial n}{\partial t}=-\frac{\epsilon^2}{2N\tau} \frac{\partial^2 n}{\partial x^2} ~~~\mathrm{if}~~~ t<t_0
\label{eqn:12}
\end{eqnarray}

This conclusion is a result of giving equal a-priori probabilities to microstates. Let us now see if the conclusion would be different if we use the new assumption instead. To do this, first note that while Equation \ref{eqn:6} gives the number of microstates of the macrostate that system is in at $t_0$, the number of microstates before K steps (at time $t_0-K\tau$) is given by

\begin{eqnarray}
(\epsilon/\delta)^N\prod_i\frac{1}{(n_i+R_{i-1}+L_{i+1}-R_i-L_i)!} \approx (\epsilon/\delta)^N\prod_i \frac{n_i^{R_i+L_i-R_{i-1}-L_{i+1}}}{n_i!}
\label{eqn:13}
\end{eqnarray}

This is because the number of particles in segment i at $t=t_0-K\tau$ is not $n_i$ but $n_i+R_{i-1}+L_{i+1}-R_i-L_i$. In order to calculate the probabilities of the states at time $t_0-K\tau$ using the assumption of equal a-priori parobabilities for macrostates at time $t_0-K\tau$, we just have to divide the previously calculated probabilities (in Equation \ref{eqn:7}) by the above number (in Equation \ref{eqn:13}) and multiply by a constant C. \\

The reason for this can be explained as follows. The probability that we are calculating (the probability for the system to have been at a particular macrostate at time $t_0-K\tau$) is the a-priori probability for that macrostate multiplied by the transition probability to the current macrostate. (This is just Bayes' theorem.) When we use our new assumption, the a-priori probabilities are decreased by a factor proportianal to the number of microstates in the macrostate at time $t_0-K\tau$ (since in the old assumption the a-priori probability of a macrostate was proporional to the number of microstates in it while in the new assumption it is independent of the number of microstates). So the probabilities that we are calculating also get divided by the same factor. \\

So for $t<t_0$, the probabilities for $R_i$ and $L_i$ to take a particular set of values according to the new assumption are

\begin{eqnarray}
P'(R_i,L_i)= \left( K!~ \prod_i  \left(\frac{n_i}{2N}\right)^{R_i+L_i} \frac{1}{ R_i!L_i! } \right) \left(C (\epsilon/\delta)^{-N}\prod_i \frac{n_i!}{n_i^{R_i+L_i-R_{i-1}-L_{i+1}}} \right) \nonumber \\ \nonumber \\
= \left( C' K!~ \prod_i  \left(\frac{n_i}{2N}\right)^{R_{i-1}+L_{i+1}} \frac{1}{ R_i!L_i! } \right)~~~~~~~~~~~~~~~~~~~~~~~~~~~~~~~~~~~~~~~~~~~
\label{eqn:14}
\end{eqnarray}

where $C'=C (\epsilon/\delta)^{-N} \prod_i n_i!$ and using the Sterling approximation this becomes

\begin{eqnarray}
\ln P'(R_i,L_i)= \ln C'+ K\ln K + \sum_i \bigg[  (R_{i-1}+L_{i+1}) \ln \left(\frac{n_i}{2N}\right) -R_i \ln R_i -L_i \ln L_i \bigg]
\label{eqn:15}
\end{eqnarray}

Using Lagrange multipliers again, this is maximum when 

\begin{eqnarray}
R_{i-1}=L_{i+1}=\frac{Kn_i}{2N}
\label{eqn:16}
\end{eqnarray}

which can also be expressed as

\begin{eqnarray}
R_i=\frac{Kn_{i+1}}{2N} ~~~\mathrm{and}~~~L_i=\frac{Kn_{i-1}}{2N}
\label{eqn:17}
\end{eqnarray}

So the change in the number of particles in segment i is 

\begin{eqnarray}
R_{i-1}+L_{i+1}-R_i-L_i=-\frac{K}{2N}(n_{i-1}+n_{i+1}-2n_i)
\label{eqn:18}
\end{eqnarray}

Note that there is an extra minus sign now (using the macrostate assumption) compared to Equation \ref{eqn:10} (which was obtained using the microstate assumption). Following the same arguments as above this gives us the correct diffusion for $t<t_0$

\begin{eqnarray}
\frac{\partial n}{\partial t}=\frac{\epsilon^2}{2N\tau} \frac{\partial^2 n}{\partial x^2} ~~~\mathrm{if}~~~ t<t_0
\label{eqn:19}
\end{eqnarray}

The above calculations show that the assigning equal a-priori probabilities to microstates gives the correct diffusion equation for $t>t_0$ but not for $t<t_0$. In order to get the correct equation for $t<t_0$ we must assign equal a-priori probabilities to macrostates instead. \\

As in the previous section, it must be noted here also that we have not used the new assumption in an accurate way since we assigned eqaul a-priori probabilities to macrostates at time $t_0-K\tau$ whereas it is macrostates at the beginning of the universe that must be assigned equal a-priori probabilities. So the above is not a rigorous proof but simply a way of getting an idea of how the new assumption works.

\section{Time and the flow of information}
Imagine that we find two books with the exact same things written on them. We will deduce that they must be two copies of something written by the same author. It would be very unlikely that two different people wrote the same thing by coincidence. Or imagine that we find someone's fingerprint on a doorpost. We will deduce that it must have been ``copied'' from his finger to the doorpost when he touched it. It would be almost impossible for the fingerprint to have appeared there by coincidence due to random collisions of air molecules and dust particles. Similarly when we observe the same DNA sequences in various living beings, we deduce that this is not by coincidence but because they inherited the DNA from a common ancestor that they evolved from. \\

\begin{SCfigure}[][h!]
\centering
\includegraphics[scale=0.4]{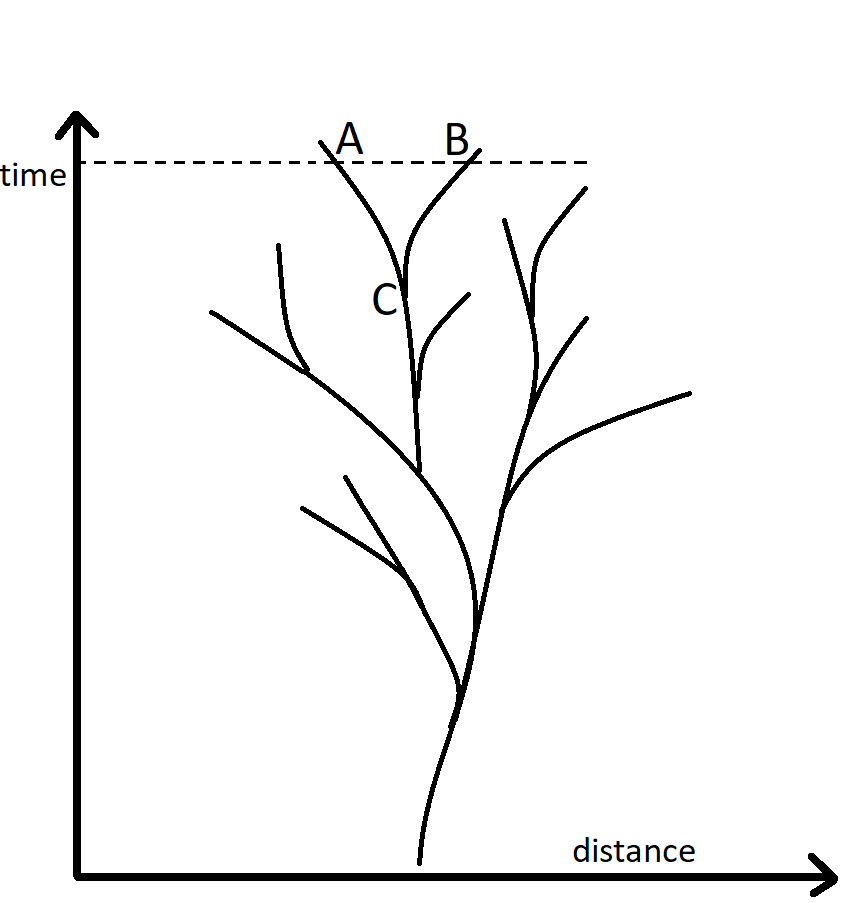} 
\caption{A representation of the flow of information. Branching out of the lines represents copying of information while the endpoints represent loss of information. \\ $~$ \\ Here, A and B represent two copies of the information at a particular time. The event at which one was copied from the other (denoted by C) is always at an earlier time.}
\label{fig:4}
\end{SCfigure}

What these examples have in common is that two identical pieces of information, when traced back to the past, are usually found to have emerged from a common source - either one was copied from the other or they were both copied from a third one. Our memories of an event can also be thought of as information copied from the actual event. But when we trace the paths of the pieces of information forward into the future, they will generally not merge together. Instead they might either generate more copies or disintegrate and become lost. This time-asymmetric flow of information, shown in Figure \ref{fig:4} forms an important part of our everyday lives. Can this also be explained in terms of the way a-priori probabilities are assigned? \\

In order to find out, let us use a simple model of the storage and copying of information. Consider a device capable of storing n bits of information. Then the information on this device could take $N=2^n$ possible values (since each bit can take 2 possible values). Every possible piece of information that can be stored on this device would then correspond to an integer between 1 and N. Let the initial state of this device be denoted by the integer $a$ with $1\le a \le N$. Consider the process of copying this information to another identical device. Let the initial state of the second device be $b$ so that the combined state of both devices could be denoted by $(a,b)$. So the copying process can be denoted by

\begin{eqnarray}
(a,b)\rightarrow(a,a)
\label{eqn:20}
\end{eqnarray}

Note that during the copying, the initial information in the second device (i.e. $b$) is lost. This leads us to an important concept called Landauer's principle \cite{landauer1961irreversibility} which states that every time information is copied from one device to another, the entropy of the surroundings must increase. To understand the reason for this, we must first understand the unitarity principle \cite{binney2013physics} which can be expressed as follows: \\

The equations of motion of an isolated system (or of the universe) specify which initial states of the system evolve into which final states. For example, if the system is initially in a state $i$ it would evolve into a final state $f$ after a certain amount of time. And if it were in another initial state $i'$ it would evolve into a final state $f'$. The unitarity principle states that if $i$ and $i'$ are different then $f$ and $f'$ must also be different. That is, two different initial states cannot evolve into the same final state. (Note that I am expressing a simplified classical version of the unitarity principle. The complete version also deals with quantum superpositions of different initial and final states). \\

Let us apply this principle to the copying process described above. The copying cannot take place in isolation from the surroundings because if it did, two different initial states $(a,b_1)$ and $(a,b_2)$ would evolve into the same final state $(a,a)$. The only way it could take place is if the state of the surroundings also changes in the process. Let the initial microstate of the surroundings be denoted by c. The final state of the surroundings would depend on the information that was already on the second device before the copying (i.e. $b_1$ or $b_2$). This can be expressed as

\begin{eqnarray}
(a,b_1,c)\rightarrow(a,a,c'_1) \nonumber \\ \nonumber \\
(a,b_2,c)\rightarrow(a,a,c'_2)
\label{eqn:21}
\end{eqnarray}

where $c'_1$ and $c'_2$ are two possible microstates in the final macrostate of the surroundings. Note that each microstate in the initial macrostate of the system can evolve into N possible microstates in the its final macrostate (depending on which of the N possible values the information was initially on the second device). So the entropy of the surroundings must increase by at least $k_B\ln N$ i.e. $nk_B\ln 2$. In other words, the surroundings must absorb at least $nk_B T_S\ln 2$ units of heat, where $T_S$ is the temperature of the surroundings. This is Landauer's principle \cite{landauer1961irreversibility}.  \\

For simplicity let us assume that during the copying process, the two devices are in contact with only a small part of the surroundings (let us call it the heat bath) which has two macrostates, the first of which has one microstate $c$ and the second has N microstates $c'_1, c'_2, \cdots c'_N$. So if the heat bath is initially in the low entropy macrostate ($c$), the copying process will always happen correctly and the heat bath will end up in the high entropy macrostate $(c'_1, c'_2, \cdots c'_N)$. What will happen if the heat bath is already in the high entropy macrostate? \\

In this case the initial state will be of the form $(a,b,c')$. Since $a$, $b$ and $c'$ can take N values each, there are $N^3$ possible initial microstates. So the unitarity principle implies that these must evolve into $N^3$ different final microstates. But states in which the information has been correctly copied must be of the form $(a,a,c')$, that is a total of $N^2$ microstates (if the heat bath is in the higher entropy macrostate) or $(a,a,c)$, that is a total of $N$ microstates (if the heat bath is in the lower one). But the $N^2$ microstates of the form $(a,a,c')$ are the ones that the system would evolve into if the heat bath was initially in the low entropy macrostate (See Equation \ref{eqn:21}). So the unitarity principle implies that if the heat bath is initially in the higher entropy macrostate, it cannot eveolve into any of these. \\

So if the heat bath is initially in the higher entropy macrostate, there are only $N$ microstates it can evolve into in which the information is correctly copied. This means that there is a probability of only $N/N^3=1/N^2$ that the information will be correctly copied. Now let us calculate the overall probability for the information to be correctly copied. \\

If we use equal a-priori probabilities for microstates, the lower and higher entropy macrostates of the heat bath will have probabilities of $1/(N+1)$ and $N/(N+1)$ respectively. As seen above, the probabilities for correct copying of the information when the heat bath is in these two macrostates are $1$ and $1/N^2$ respectively. So the overall probability for correct copying is

\begin{eqnarray}
\frac{1}{N+1}\times 1+\frac{N}{N+1}\times \frac{1}{N^2}=\frac{1}{N}
\label{eqn:22}
\end{eqnarray}

which is negligibly small (because $N=2^n$ is very big). But if we use  equal a-priori probabilities for macrostates, then the two macrostates of the heat bath would have probabilities of $1/2$ each. Hence the probability for correct copying would be

\begin{eqnarray}
\frac{1}{2}\times 1+\frac{1}{2}\times \frac{1}{N^2} \approx \frac{1}{2}
\label{eqn:23}
\end{eqnarray}

Let us apply the above results to the problem discussed at the beginning of this section. That is, given two identical pieces of information, comparing the probabilities that they were copied from a common source and the probability that they happen to be the same by random chance. For this let us consider a system like the one studied above (consisting of two storage devices and a heat bath). Assume that one of the following two processes is performed on it:

\begin{itemize}
\item With probability p, the information on the first device is copied to the second by the process described above or
\item with a probability $(1-p)$ the information on the second device is changed to a random integer between 1 and N.
\end{itemize}

After one of these processes is done (but we don't know which one), suppose we find that the two devices contain the same information. What are the probabilities for this to have happened as a result of correct copying (the first process) vs as a result of a random change (the second process)? \\

First let us calculate using equal a-priori probabilities for microstates. In this case, the probability for correct copying is $1/N$ (Equation \ref{eqn:22}). The probability for a random change to result in the same information is also $1/N$ because the information stored on the device can take N possible values. So Bayes' theorem implies that the probability that it was copied is 

\begin{eqnarray}
\frac{p\times (1/N)}{p\times (1/N)+(1-p)\times (1/N)}=p
\label{eqn:24}
\end{eqnarray}

and the probability that it came up randomly is

\begin{eqnarray}
\frac{(1-p)\times (1/N)}{p\times (1/N)+(1-p)\times (1/N)}=1-p
\label{eqn:25}
\end{eqnarray}

This answer is strongly inconsistent with our experiences and our intuitive sense of probabilities. To see why, consider the case of $p=1/2$ and $n=100$ (so that $N\approx 10^{30}$). Since the 100 bits on the second device together can take $10^{30}$ possible values, if they are identical to the 100 bits on the first device, we would deduce they were almost certainly copied. But the above calculation gives a probability of $1/2$ for them to have been copied and $1/2$ for them to have become identical to the ones on the first device by random chance. \\

So let us try again using equal a-priori probabilities for macrostates instead. Now the probability for correct copying is $1/2$ (Equation \ref{eqn:23}). So using Bayes' theorem again, the probability that the information was copied is

\begin{eqnarray}
\frac{p\times (1/2)}{p\times (1/2)+(1-p)\times (1/N)}
\label{eqn:26}
\end{eqnarray}

which approaches 1 for large N and the probability that it came up randomly is

\begin{eqnarray}
\frac{(1-p)\times (1/N)}{p\times (1/2)+(1-p)\times (1/N)}
\label{eqn:27}
\end{eqnarray}

which approaches zero for large N. Now the probabilities are consistent with our intuition. \\

The ``information storage devices'' discussed above do not necessarily have to be computer chips. They could be any of the variety of ways information is stored in nature, such as genetic information in DNA, the things we have read or heard stored in our brains, information in fossils about ancient life forms etc. Each time we come across such a piece of information we do not assume that it is something that just appeared randomly. If we do, we would have to conclude that all information we see is meaningless and we would also have no explanation for why nature is filled with multiple copies of the same information - such as the DNA in each of the trillions of cells in each of the trillions of living beings on the earth. \\

Instead we assume that it was copied - perhaps with some errors - from a source. Such an assumption seems natural and intuitive to us but the calculations above show that this assumption is not consistent with equal a-priori probabilities for microstates. Equal a-priori probabilities for macrostates at an early time, on the other hand, is consistent with our intuition.

\section{Some speculations on the reason for the new assumption}
What could be the reason for assigning equal a-priori probabilities to macrostates at the beginning of the universe? Even though this assumption explains several observations better than the usual assumption of equal a-priori probabilities for microstates, it doesn't seem right at first for the following reason. Suppose there are $W_{tot}$ microstates that are classified into V macrostates, with $W_M$ microstates in macrostate M (with $\sum_M W_M = W_{tot}$) This means that in our new assumption, the probability assigned to macrostate M is $1/V$ and the average probability assigned to each microstate within that macrostate is $1/(VW_M)$. \\

This probability depends on $W_M$ which is the number of microstates in the same macrostate, or in other words, the number of microstates that an observer cannot distinguish from the given microstate. How can something fundamental like the a-priori probabilities depend on what an observer can measure or distinguish? Below, I will explore some reasons why this could be the case. These are just speculations. None of these reasons are a definite proof that these are the correct a-priori probabilities. \\

Suppose a particular macrostate M contains the microstates $m$ and $m'$ (among others). This means the observer will get the same result for measurements performed on the system whether it is in $m$ or $m'$. But it is possible that when the system evolves with time, it might evolve into two different macrostates depending on whether it was initially in $m$ or $m'$. So even though $m$ and $m'$ belong to the same macrostate, they can still be distinguishable because of the way the system evolves. So for simplicity, I will first consider only non-evolving systems. (This won't be an accurate model as almost everything around us is evolving with time.) Then later, I will try to take the evolution of the system into account. \\

Let us consider a simple model consisting of only an observer and a system whose microstate does not change. Assume that the system has V macrostates with $W_M$ microstates in macrostate M. Any two microstates that are in the same macrostate will interact with the observer in the same way during a measurement. Since for this non-evolving system, microstates within a macrostate are completely indistinguishable, this suggests that those microstates are, in a way, the same and thus the a-priori probabilty of the macrostate should not depend on how many microstates there are. That is, they do not needed to be counted separately. \\

The above argument is certainly not a rigorous one but let us see if there is anything else that suggests that it may be correct. For this, I will use the derivation of the law of equal a-priori probabilities using the principle of maximum information \cite{jaynes1957information, jaynes1957information2}. This derivation starts from the question, if we know only the possible microstates a system can be in, and we have no other information, what a-priori probabilities can we assign them? One reasonable way to do this is to consider the amount of information that we would obtain if we found out which microstate the system is in. (A way of quantifying information has been provided by Shannon \cite{shannon1948mathematical}. If we get a result to which we had previously assigned a probability of p, the amount of information gained is $-\ln p$.) Using this, if we assign probability $p_m$ to microstate m then the average amount of information we will get when we find out which microstate the system is in is

\begin{eqnarray}
-\sum_{m=1}^W p_m \ln p_m
\label{eqn:28}
\end{eqnarray}

Here I am using lowercase indices like m to range over all microstates (from 1 to W) and uppercase indices like M to range over all macrostates (from 1 to V). Using the method of Lagrange multipliers \cite{paulsonlinenotes:2019}, it can be shown that this quantity is maximum when all the probabilities are equal. Should this calculation be modified when some microstates (those belonging to the same macrostate) are indistinguishable from others? In this case, if the system is in microstate m, a measurement will not confirm this. The measurement will only tell us that the system is in the macrostate M to which m belongs, i.e. it could be in any of the $W_M$ microstates in M. So the amount of information gained by this measurement is not $-\ln p_m$ but $-\ln \sum_{n\sim m} p_n $ where the subscript $n\sim m$ on the sum denotes that the sum is over all microstates n that belong to the same macrostate as m. So the average amount of information resulting from the measurement is 

\begin{eqnarray}
I=-\sum_{m=1}^W p_m \ln \sum_{n\sim m} p_n=-\sum_{m=1}^W p_m \ln p_M
\label{eqn:29}
\end{eqnarray}

where $p_M=\sum_{n\sim m} p_n$ is the total probability of all microstates in the same macrostate M as microstate m. As I show in Appendix 1, this quantity takes the maximum possible value when all \emph{macrostates} are given equal a-priori probabilities. \\

Let us now try to use a different argument that suggests that macrostates should have equal a-priori probabilities. Consider two macrostates A and B in our simple model, with 7 and 3 microatates respectively. It seems intuitive to us that their probabilities should be in the ratio 7:3, just as if we had 7 green marbles and 3 red marbles in a box, the probability of randomly taking a green or red one out of the box would be in the ratio 7:3 \\

Could this have anything to do with the marbles being distinguishable (because they won't be exactly identical) whereas in our model of a non-evolving system, the microstates within a macrostate are assumed to be identical? To avoid this confusion, let's imagine we have 7 exactly identical green marbles and 3 exactly identical red ones in a box. Now after we have taken out one marble, there would be 9 left in 9 different locations within the box. If we had taken out a red one, then out of the 9 locations, 7 should contain green marbles and 2 should contain red ones. So using combinatorics, the number of arrangements in which this can happen is $9!/(7!~2!)$. Similarly if we had taken out a green marble, the remaining 6 green and 3 red marbles could be arranged in $9!/(6!~3!)$ ways in their 9 locations in the box. The ratios of the numbers of possible arrangements is 

\begin{eqnarray}
\frac{9!}{6!~3!}:\frac{9!}{7!~2!}=7:3
\label{eqn:30}
\end{eqnarray}

which is just the ratio of the probabilities. So one way of thinking about this is that we have a higher probability of picking a green marble because then there will be more ways of arranging the remaining marbles. \\

Now a microstate is not a physical object like a marble and so it doesn't make sense to talk of its ``location.'' So when we choose a microstate from either macrostate A or B, the number of ways of arranging the remaining microstates will not be $9!/(6!~3!)$ and $9!/(7!~2!)$ respectively but just 1 and 1. So the ratios of their probabilities must be 1:1. \\

(But as mentioned earlier, these calculations for a non-evolving system are not a good model of the real world. So in Appendix 2, I will try find out whether the new assumption is also justified in the case of a system that evolves with time.) \\

\section*{Appendix 1: Maximising the amount of information gained when we find out which macrostate a system is in}

To maximise the quantity

\begin{eqnarray}
I=-\sum_{m=1}^W p_m \ln \sum_{n\sim m} p_n=-\sum_{m=1}^W p_m \ln p_M
\label{eqn:31}
\end{eqnarray}

using Lagrange multipliers, let us first calculate the derivative

\begin{eqnarray}
\frac{\partial p_M}{\partial p_k}=\frac{\partial}{\partial p_k}\sum_{n\sim m} p_n=\sum_{n\sim m} \delta_{nk}=1~~~\mathrm{if}~~~k\sim m ~~~\mathrm{and}~~~0~~~\mathrm{otherwise}
\label{eqn:32}
\end{eqnarray}

Hence

\begin{eqnarray}
\frac{\partial \ln p_M}{\partial p_k}=\frac{1}{p_M}~~~\mathrm{if}~~~k\sim m ~~~\mathrm{and}~~~0~~~\mathrm{otherwise}
\label{eqn:33}
\end{eqnarray}

Using this we can find the derivative of the average information with respect to one of the probabilities.

\begin{eqnarray}
\frac{\partial I}{\partial p_k}=-\sum_m \frac{\partial p_m}{\partial p_k}  \ln p_M - \sum_m p_m \frac{\partial \ln p_M}{\partial p_k} \nonumber \\ \nonumber \\ \nonumber \\
=-\sum_m \delta_{km} \ln p_M -\sum_{m\sim k} p_m  \frac{1}{p_M} ~~~~~~~~~~~~~ \nonumber \\ \nonumber \\
=-\ln p_K -1 ~~~~~~~~~~~~~~~~~~~~~~~~~ 
\label{eqn:34}
\end{eqnarray}

Since the constraint is $\sum_i p_i=1$ and $(\partial/\partial p_k) \sum_i p_i= 1$ this means $\partial I/\partial p_k$ must be equal to the Lagrange multiplier $\lambda$. This means

\begin{eqnarray}
-\ln p_K -1=\lambda ~~~~~~~~~~~~~~ \nonumber \\ \nonumber \\
\therefore p_K=e^{-\lambda-1} ~~~\mathrm{for~all~K} 
\label{eqn:35}
\end{eqnarray}

In other words, all macrostates have the same equal a-priori probabilities.

\section*{Appendix 2: Speculations on the reason for the new assumption for an evolving system}
In Section 5, I considered the case where the system being observed remains unchanged in the same microstate. Here, I will remove this restriction and see if our new assumption is still justified. I will now assume that out of the 7 microstates in macrostate A, one evolves into a microstate in macrostate B after a certain amount of time $t_0$ while 6 remain in macrostate A. Out of the 3 microstates in macrostate B, one evolves into a microstate in macrostate A while 2 remain in  macrostate B. \\

To make the calculation simpler I will also assume that the observer is present throughout the evolution of the system. This is not an accurate assumption because the initial state has to be at the beginning of the universe (and the new assumption is that macrostates have equal a-priori probabilities at the beginning of the universe) but we know that life forms emerged only long after the universe formed. But because of the difficulty in modelling a complex process such as the origin of life, I will ignore this problem. \\


At $t=t_0$ the observer will be able to find out the current macrostate of the system and will also remember its initial macrostate at $t=0$. So there are 4 possibilities: one is that the system could have been in macrostate A initially and could still be in macrostate A now - I will denote this as $(A\rightarrow A)$. Similarly there are 3 other possibilities that can be denoted by $(A\rightarrow B)$, $(B\rightarrow A)$ and $(B\rightarrow B)$ respectively. If we use equal a-priori probabilities for microstates we get

\begin{eqnarray}
P(A\rightarrow A)=6/10  \nonumber \\ \nonumber \\
P(A\rightarrow B)=1/10  \nonumber \\ \nonumber \\
P(B\rightarrow A)=1/10  \nonumber \\ \nonumber \\
P(B\rightarrow B)=2/10
\label{eqn:36}
\end{eqnarray}

i.e. the probabilities are just the number of microstates corresponding to each case divided by the total number of microstates which is 10. \\

If we use the new assumption instead then the sum of probabilities $P'(A\rightarrow A)+P'(A\rightarrow B)=1/2$ and $P'(B\rightarrow A)+P'(B\rightarrow B)=1/2$. (Because the probabilities of starting out in macrostates A or B must be 1/2 each.) Since 6 of the 7 microstates of A remain in A while one goes into B this means 

\begin{eqnarray}
P'(A\rightarrow A)=6/7 \times 1/2  \nonumber \\ \nonumber \\
P'(A\rightarrow B)=1/7 \times 1/2
\label{eqn:37}
\end{eqnarray}

Similarly 

\begin{eqnarray}
P'(B\rightarrow A)=1/3 \times 1/2  \nonumber \\ \nonumber \\
P'(B\rightarrow B)=2/3 \times 1/2
\label{eqn:38}
\end{eqnarray}

Below I will present an argument for why these could be the correct probabilities. This argument is not rigorous and might be flawed, but since our new assumption explains so many of our observations that cannot be explained using the old (microstate) assumption, it is important to explore the possible reasons behind it.

\begin{itemize}

\item Consider the case $(A\rightarrow A)$. The initial microstate of the system was in A and its present microstate is also in A. Other than the initial microstate, there are 6 more in A, of which 1 must go to B and 5 must remain in A. The number of ways in which this can happen is $6!/(5!~1!)$ \\


\emph{In this step I have made the assumption that one microstate (the one that the system is initially in) is not included when calculating the number of ways the different microstates can evolve into one another. This is why we get $6!/(5!~1!)$ as the answer and not of $7!/(6!~1!)$. There is some justification for this as there is a fundamental difference between the microstate that we are excluding (it is the true microstate that the universe began in) and the other microstates (they are just hypothethical microstates that the universe could have begun in). Still the validity of this assumption needs to be explored further.}

\item Similarly for $(A\rightarrow B)$, the initial microstate of the system was in A and its present microstate is in B. Other than the initial microstate, there are 6 more in A, all of which must remain in A. The number of ways in which this can happen is $6!/(6!~0!)$ \\

\item For both of the above cases, there would also be 3 microstates in B of which 2 must remain in B and 1 must go to A. The number of ways for this is $3!/(2!~1!)$. \\

\item And for $(B\rightarrow A)$, the initial microstate of the system was in B and its present microstate is in A. Other than the initial microstate, there are 2 more in B, all of which must remain in B. The number of ways in which this can happen is $2!/(2!~0!)$ \\

\item Finally for $(B\rightarrow B)$, the initial microstate of the system was in B and its present microstate is also in B. Other than the initial microstate, there are 2 more in B, 1 of which must go to A and 1 must remain in B. The number of ways in which this can happen is $2!/(1!~1!)$ \\

\item For the last two cases, there would also be 7 microstates in A of which 6 must remain in A and 1 must go to B. The number of ways for this is $7!/(6!~1!)$. \\

\end{itemize}

Assuming that the probability is proportional to the number of ways of the remaining microstates (other than the true initial microstate) can evolve into one another, the ratio of probabilities is:

\begin{eqnarray}
P'(A\rightarrow A)~:~P'(A\rightarrow B)~:~P'(B\rightarrow A)~:~P'(B\rightarrow B) ~~~~~~~~~~~~~~~~~~~~ \nonumber \\ \nonumber \\
=\left(\frac{6!}{5!~1!}\right) \left(\frac{3!}{2!~1!}\right)~:~\left(\frac{6!}{6!~0!}\right) \left(\frac{3!}{2!~1!}\right)~:~\left(\frac{2!}{2!~0!}\right) \left(\frac{7!}{6!~1!}\right)~:~\left(\frac{2!}{1!~1!}\right) \left(\frac{7!}{6!~1!}\right) \nonumber \\ \nonumber \\
=\frac{6}{7}~:~\frac{1}{7}~:~\frac{1}{3}~:~\frac{2}{3}  ~~~~~~~~~~~~~~~~~~~~~~~~~~~~~~~~~~~~~~~~~
\label{eqn:39}
\end{eqnarray}

which means we get the correct probabilities as required. Note also that while I have used specific numbers in this example (7 and 3 microstates in each macrostate with 1 from each going to the other macrostate) the same principle applies if we use any other numbers and we would still get the correct probabilities. \\

But there are several problems with this reasoning. First, as mentioned above, this method uses the number of ways the microstates \emph{other than the initial one} at $t=0$ can evolve into the microstates \emph{other than the present one} at $t=t_0$. But we are ignoring the possibility of a different microstate evolving into the present microstate, that is, we are ignoring the possibility that a different microstate could have been the initial microstate. It is not clear whether doing the calculation this way is valid. \\

Secondly, for example in the $(A\rightarrow A)$ case, where out of the 6 remaining microstates 1 must go to B, we assumed that we don't know which of the 6 microstates it is and included all the possibilities. That is how we got the factor of $6!/(5!~1!)$. But even though the observer cannot distinguish between these 6 microstates, they are not the same. So there is only one of them that can go to B. This seems to imply that considering all 6 possibilities is not correct. This also needs to be explored further. \\

Third, this analysis used a classical model. To be accurate, we would need a quantum model in which quantum superpositions of the various microstates are also considered. \\

\medskip
\bibliography{ms}
\end{document}